\documentclass[aps,prl,twocolumn,floats,epsfig,showpacs]{revtex4-1}

\usepackage{bbm}
\usepackage{amssymb}
\usepackage{ifpdf}
\usepackage{color}
\usepackage{graphicx,epsfig}

\newcommand{\Z}{{\mathbb{Z}}}

\newcommand{\RP}{{\mathbb{R}P}}

\begin{document}

\title{Interfaces, Strings, and a Soft Mode in the Square Lattice Quantum Dimer 
Model}
\author{D.\ Banerjee$^1$, M.\ B\"ogli$^1$, C.\ P.\ Hofmann$^2$, 
F.-J.\ Jiang$^3$, P.\ Widmer$^1$, and U.-J.\ Wiese$^1$}
\affiliation{$^1$Albert Einstein Center, Institute for Theoretical Physics, 
Bern University, Switzerland \\
$^2$ Facultad de Ciencias, Universidad de Colima, Bernal Diaz del Castillo 340,
Colima C.P.\ 28045, Mexico \\
$^3$ Department of Physics, National Taiwan Normal University
88, Sec.\ 4, Ting-Chou Rd., Taipei 116, Taiwan}

\begin{abstract}

The quantum dimer model on the square lattice is equivalent to a $U(1)$ gauge 
theory. Quantum Monte Carlo calculations reveal that, for values of the 
Rokhsar-Kivelson (RK) coupling $\lambda < 1$, the theory exists in a 
confining columnar phase. The interfaces separating distinct columnar phases 
display plaquette order, which, however, is not realized as a bulk phase. Static
``electric'' charges are confined by flux tubes that consist of multiple 
strands, each carrying a fractionalized flux $\frac{1}{4}$. A soft
pseudo-Goldstone mode emerges around $\lambda \approx 0$, long before one 
reaches the RK point at $\lambda = 1$.

\end{abstract}

\maketitle

Quantum dimer models \cite{Rok88} implement Anderson's ideas of resonating 
valence bonds \cite{And87} as a potential route towards understanding 
high-temperature superconductivity. A dimer connecting neighboring lattice sites
represents a singlet-pair of two spins $\frac{1}{2}$. In the undoped dimer model
each lattice site is touched by exactly one dimer, thus modeling a system in 
which each spin participates in a singlet pair. A lot of progress has been made 
on unraveling the phase structure of classical and quantum dimer models for 
various lattice geometries 
\cite{Sac89,Lev90,Leu96,Moe02,Hen04,Ale05,Ale06,Cha10,Can10,Alb10,Tan11,Lam13}. 
Despite the fact that they do not suffer from the notorious sign problem, 
quantum Monte Carlo calculations for quantum dimer models 
are rather limited. For the square lattice, Green's function Monte Carlo 
calculations \cite{Syl05} have so far led to inconclusive results. While 
\cite{Syl06} found evidence for a phase transition between the columnar and the 
plaquette phase (cf.\ Fig.1a,b) near $\lambda \approx 0.6$, \cite{Ral08} 
concluded that for $\lambda \gtrsim 0$ there might be a mixed phase that shares 
features of both columnar and plaquette phases.

The square lattice quantum dimer model is closely related to the $(2+1)$-d 
$U(1)$ quantum link model \cite{Hor81,Orl90,Cha97}, a lattice gauge theory with
a 2-dimensional spin $\frac{1}{2}$ Hilbert space per link, which has also been
investigated in the context of spin liquids \cite{Her04,Sha04}. In fact, both 
models share the same Hamiltonian, but realize the Gauss law in two different 
ways. While the quantum link model has no background charges in its ground 
state, the quantum dimer model operates in a staggered background of charges 
$\pm 1$, which ensures that each lattice site is touched by exactly one dimer. 
Quantum link models \cite{Cha97,Bro99,Bro04} provide an alternative 
non-perturbative regularization of gauge theories in particle physics. In the 
quantum link formulation of 4-d Quantum Chromodynamics, the confining gluon 
field emerges by dimensional reduction from a deconfined Coulomb phase of a 
$(4+1)$-d $SU(3)$ quantum link model, while chiral quarks arise naturally as 
domain wall fermions located at the two 4-d sides of a $(4+1)$-d slab
\cite{Bro99}. Thanks to their finite-dimensional link Hilbert space, quantum 
link models are ideally suited for the constructions of atomic quantum 
simulators \cite{Wie13} for Abelian \cite{Bue05,Mue09,Mue10,Zoh12,Ban12} and 
non-Abelian gauge theories \cite{Tag12,Tag12a,Ban13} using ultracold atoms in 
an optical lattice.
\begin{figure}[tbp]
\includegraphics[width=0.46\textwidth]{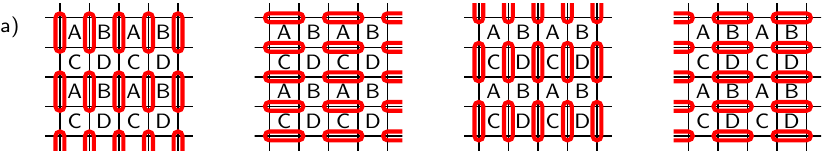} \\ \vskip0.1cm
\includegraphics[width=0.46\textwidth]{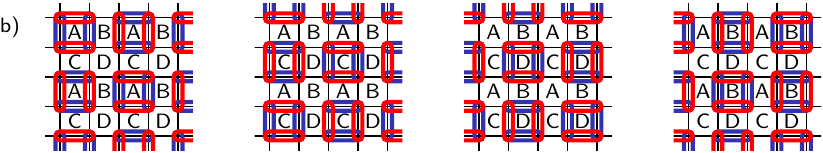} \\
\includegraphics[width=0.46\textwidth]{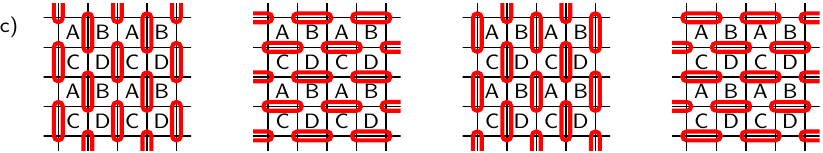} \\
\caption{[Color online] \textit{a) Columnar, b) plaquette, and c) staggered 
order, shown with four dual sublattices $A$, $B$, $C$, $D$.}}
\end{figure}

Recently, we have developed a very efficient cluster algorithm to simulate the
$(2+1)$-d $U(1)$ quantum link model, which has allowed us to study its confining
dynamics \cite{Ban13a}. In particular, there are two distinct confining phases
(analogous to columnar and plaquette phases in the quantum dimer model)
with different discrete symmetry breaking patterns, separated by a weak first 
order phase transition that mimics several features of deconfined quantum 
critical points \cite{Vis04,Sen04,Sen04a}. In particular, at the phase 
transition a light pseudo-Goldstone
boson emerges dynamically, which can, however, not be interpreted as a dual
photon, because it is not exactly massless. The confining strings that connect 
an external charge-anti-charge pair fractionalize into mutually repelling 
strands, each carrying fractional flux $\frac{1}{2}$. The interior of the 
strands consists of the bulk phase that is stable on the other side of the
phase transition. 

Here we apply numerical simulations to the square lattice 
quantum dimer model. We find that, in the columnar phase, flux strings
fractionalize into strands of flux $\frac{1}{4}$, whose interior consists of 
plaquette phase, which, however, is not realized in the bulk. The same is true 
for the interfaces separating distinct columnar phases, which display complete 
wetting. Unlike in the quantum link model, in the quantum dimer model there is 
no phase transition separating two distinct confining phases. Instead the 
columnar phase extends all the way to the RK point $\lambda = 1$. Remarkably, 
on moderate volumes the massless mode that arises at the RK point influences 
the entire region $0 \lesssim \lambda < 1$, in which it is still relatively 
light.

The configurations of the quantum dimer model are characterized by variables
$D_{xy} \in \{0,1\}$, indicating the presence or absence of a dimer on the link
connecting the neighboring sites $x$ and $y$ on a square lattice. The 
Hamiltonian of the quantum dimer model is the same as the one of the $(2+1)$-d 
$U(1)$ quantum link model
\begin{equation}
H = - J \sum_{\Box} \left[U_\Box + U_\Box^\dagger - 
\lambda (U_\Box + U_\Box^\dagger)^2\right].
\end{equation}
Here $U_\Box = U_{wx} U_{xy} U_{zy}^\dagger U_{wz}^\dagger$ is a plaquette operator 
formed by quantum links $U_{xy}$ connecting nearest-neighbor sites $x$ and $y$. 
A $U(1)$ quantum link $U_{xy} = S_{xy}^+$ is a raising 
operator of electric flux $E_{xy} = S_{xy}^3$, constructed from a quantum spin 
$\frac{1}{2}$ associated with the link $xy$. As shown in Fig.2, the electric 
flux variables are related to the dimer variables by 
$E_{xy} = (-1)^{x_1+x_2} (D_{xy} - \frac{1}{2})$.
\begin{figure}[tbp]
\includegraphics[width=0.26\textwidth]{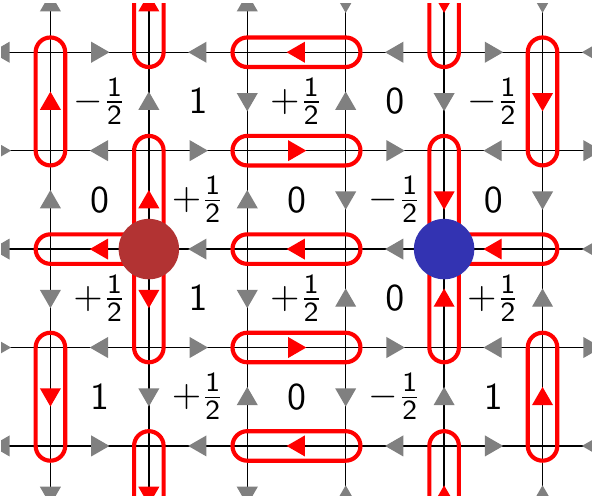}
\caption{[Color online] \textit{Dimer configuration with associated fluxes and
dual height variables, in the presence of two external charges.}}
\end{figure}
The first term in the Hamiltonian rotates a pair of parallel dimers on opposite
links of a plaquette by 90 degrees. Equivalently, it flips a loop of electric 
flux, winding around the plaquette. It also annihilates non-flippable plaquette 
states, while the RK term, proportional to $\lambda$, counts flippable 
plaquettes. The Hamiltonian commutes with the generators 
$G_x = \sum_i (E_{x,x+\hat i} - E_{x-\hat i,x})$ of infinitesimal $U(1)$ gauge 
transformations. In allowed configurations exactly one dimer touches each site, 
such that
\begin{equation}
G_x = (-1)^{x_1+x_2} \sum_i (D_{x,x+\hat i} + D_{x-\hat i,x}) = (-1)^{x_1+x_2},
\end{equation}
where $\hat i$ is the unit-vector in the $i$-direction. The dimer covering
constraint is thus equivalent to a staggered background of electric charges 
$\pm 1$. While in the quantum link model physical states obey 
$G_x |\Psi\rangle = 0$, in the quantum dimer model they obey 
$G_x |\Psi\rangle = (-1)^{x_1+x_2} |\Psi\rangle$. 

Besides the $U(1)$ gauge symmetry, the model also has several global symmetries,
including translations by one lattice spacing combined with charge conjugation,
CT$_x$ and CT$_y$ (which correspond to ordinary translations of the dimers 
$D_{xy}$), a 90 degrees rotation O around a lattice point, as well as a
90 degrees rotation around a plaquette center combined with charge conjugation
CO$'$. Another important global symmetry is the $U(1)^2$ center symmetry 
associated with ``large'' gauge transformations \cite{tHo79}. On an 
$L_1 \times L_2$ lattice with periodic boundary conditions, there are 
super-selection sectors characterized by wrapping electric fluxes 
$E_i = \frac{1}{L_i} \sum_x E_{x,x+\hat i} \in \Z/2$. They commute with the 
Hamiltonian, but cannot be expressed through ``small'' periodic gauge 
transformations $G_x$.

\begin{figure}[tbp]
\includegraphics[width=0.46\textwidth]{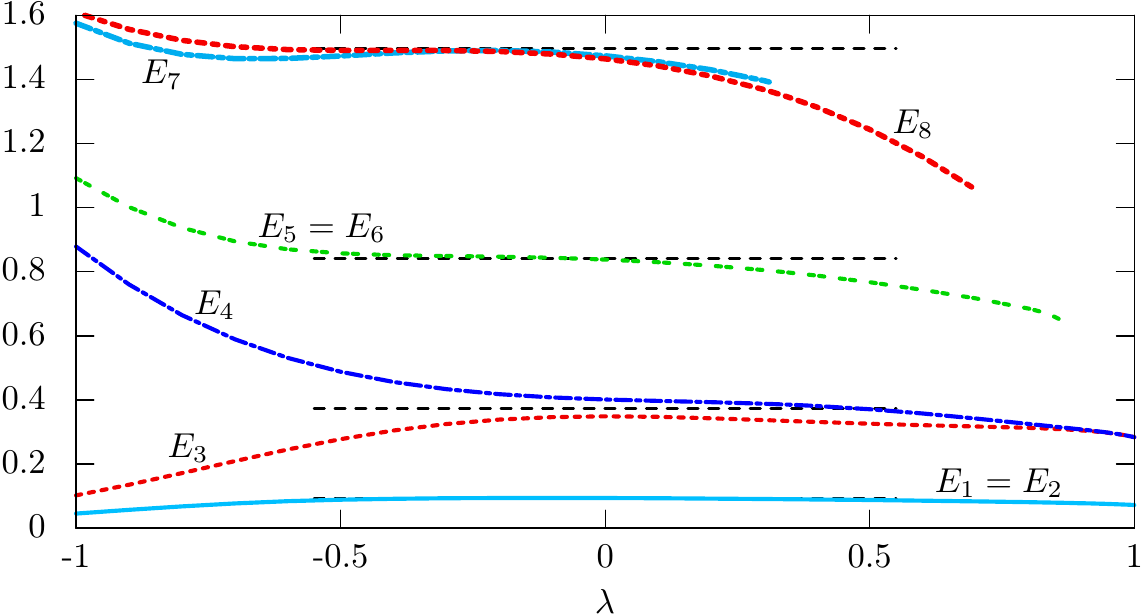}
\caption{[Color online] \textit{Energy spectrum on an $8^2$ lattice as a 
function of the RK coupling $\lambda$.}}
\end{figure}
Following \cite{Ral08}, we have performed exact diagonalization studies on 
$L_1 \times L_2$ lattices with $L_1, L_2 \in \{4,6,8\}$. The energies of some of
the lowest states are illustrated in Fig.3. For $\lambda < 1$, the ground state 
has quantum numbers (CT$_x$,CT$_y$) = (+,+). The first two excited states with
energy gap $E_1 = E_2$ are degenerate and have quantum numbers $(+,-)$ and 
$(-,+)$, while the next excited state with energy gap $E_3$ has quantum numbers 
$(+,+)$. For $\lambda \lesssim - 0.2$, the energy gaps 
$E_{1,2}, E_3 \sim \exp(- \alpha L_1 L_2)$ decrease exponentially with the volume
$L_1 L_2$, thus indicating the spontaneous breakdown of translation invariance 
that characterizes the columnar phase. For $-0.2 \lesssim \lambda \lesssim 0.8$,
a $(-,-)$ state with energy $E_4 \approx E_3$ almost degenerates with the 
$(+,+)$ state. The next exactly degenerate states with energy $E_5 = E_6$
again have quantum numbers $(+,-)$ and $(-,+)$, while the states with energies
$E_7$ and $E_8$ are almost degenerate and transform as $(+,+)$ and $(-,-)$. As
indicated by the dashed lines in Fig.3, the energy ratios of these states are 
given by $E_{1,2}:E_{3,4}:E_{5,6}:E_{7,8} \approx 1:4:9:16$, thus indicating an 
approximate rotor spectrum and possibly the transition to a different phase. 
However, the volumes accessible to exact diagonalization are too small to come 
to a definitive conclusion concerning this issue.

Green's function Monte Carlo simulations of quantum dimer models have been
performed in \cite{Syl05,Syl06,Ral08} on systems of size up to $48^2$. While 
\cite{Syl06} reached the conclusion that the columnar phase turns into the
plaquette phase near $\lambda \approx 0.6$, \cite{Ral08} interpreted the data
in terms of a mixed phase that shares features of the columnar and the 
plaquette phase. Here we apply an alternative numerical method on volumes up to 
$144^2$, and conclude that the columnar phase extends all the way to the RK 
point. First, we introduce dual height variables associated with two even 
($A$, $D$) and two odd ($B$, $C$) dual sublattices (cf.\ Fig.1), 
$h^{A,D}_{\widetilde x} = 0,1$, $h^{B,C}_{\widetilde x} = \pm \frac{1}{2}$, located at 
the dual sites $\widetilde x = (x_1 + \frac{1}{2},x_2 + \frac{1}{2})$. They are 
associated with a flux configuration $E_{x,x+\hat 1} = 
[h^X_{\widetilde x} - h^{X'}_{\widetilde x - \hat 2}] \mbox{mod} 2 = \pm \frac{1}{2}$,
$E_{x,x+\hat 2} = (-1)^{x_1+x_2} [h^X_{\widetilde x} - h^{X'}_{\widetilde x - \hat 1}] 
\mbox{mod} 2 = \pm \frac{1}{2}$, $X,X' \in \{A,B,C,D\}$. The different symmetry 
breaking patterns are distinguished by four order parameters,
$M_X = \sum_{\widetilde x \in X} s^X_{\widetilde x} h^X_{\widetilde x}$, with
$s^A_{\widetilde x} = s^C_{\widetilde x} = (-1)^{({\widetilde x}_1 + \frac{1}{2})/2}$
(${\widetilde x}_1 + \frac{1}{2}$ even),
$s^B_{\widetilde x} = s^D_{\widetilde x} = (-1)^{({\widetilde x}_1 - \frac{1}{2})/2}$
(${\widetilde x}_1 + \frac{1}{2}$ odd). The order parameters
\begin{eqnarray}
&&M_{11} = M_A - M_B - M_C + M_D = M_1 \cos\varphi_1, \nonumber \\
&&M_{22} = M_A + M_B - M_C - M_D = M_1 \sin\varphi_1, \nonumber \\
&&M_{12} = M_A - M_B - M_C - M_D = M_2 \cos\varphi_2, \nonumber \\
&&M_{21} = - M_A + M_B - M_C - M_D = M_2 \sin\varphi_2,
\end{eqnarray}
give rise to $\varphi = \frac{1}{2}(\varphi_1 + \varphi_2 + \frac{\pi}{4})$, 
which transforms as
\begin{eqnarray}
&&^{CT_x}\varphi = \pi - \varphi, \quad
^{CT_y}\varphi = \frac{\pi}{2} - \varphi, \nonumber \\
&&^O\varphi = \frac{\pi}{4} + \varphi, \quad
^{CO'}\varphi = - \frac{\pi}{4} - \varphi.
\end{eqnarray}
It should be noted that $\pm (M_A,M_B,M_C,M_D)$ (and thus $\varphi$ and 
$\varphi + \pi$) represent the same physical configuration, because shifting 
the height variables to 
$h^X_{\widetilde x}(t)' = [h^X_{\widetilde x}(t) + 1] \mbox{mod} 2$
leaves the dimer configuration unchanged. The columnar phases correspond to
$\varphi = 0 \mbox{mod} \frac{\pi}{4}$, while the plaquette phases correspond
to $\varphi = \frac{\pi}{8} \mbox{mod} \frac{\pi}{4}$.

We have performed quantum Monte Carlo 
simulations with a Metropolis algorithm applied to the dual height variables. 
The algorithm is restricted to a fixed sector of wrapping electric fluxes
(here $E_i = 0$), but updates a given sector ergodically. For $\lambda < 1$, at 
low temperatures the restriction to $E_i = 0$ is no problem, because the ground
state belongs to this sector. About three quarters of all height variable flips 
are forbidden because they violate the dimer covering constraint, and only a 
few percent of the proposed flips are accepted in the Metropolis step. Although 
the algorithm thus has a rather small acceptance rate, it works remarkably well 
and allows us to access volumes as large as $144^2$ and temperatures as low as 
$T = J/500$. The algorithm has been used to determine the probability 
distribution $p(M_{11},M_{22})$ shown in Fig.4a,b at $\lambda = - 0.5$ and $0.8$ 
for $L_1 = L_2 = 24 a$, which reveals an emergent approximate spontaneously 
broken $SO(2)$ symmetry for $\lambda \gtrsim - 0.2$.
\begin{figure}[tbp]
\includegraphics[width=0.235\textwidth]{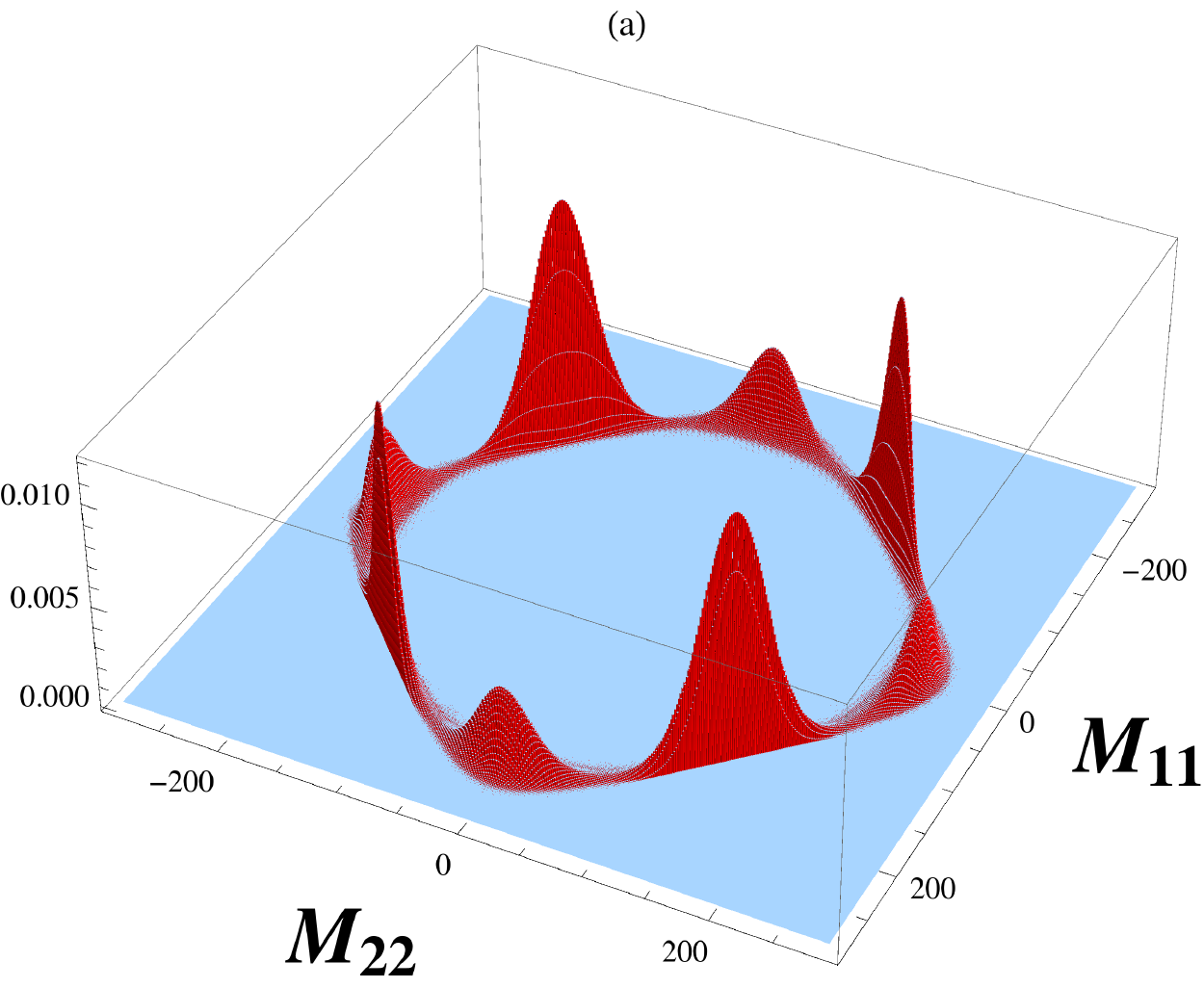}
\includegraphics[width=0.235\textwidth]{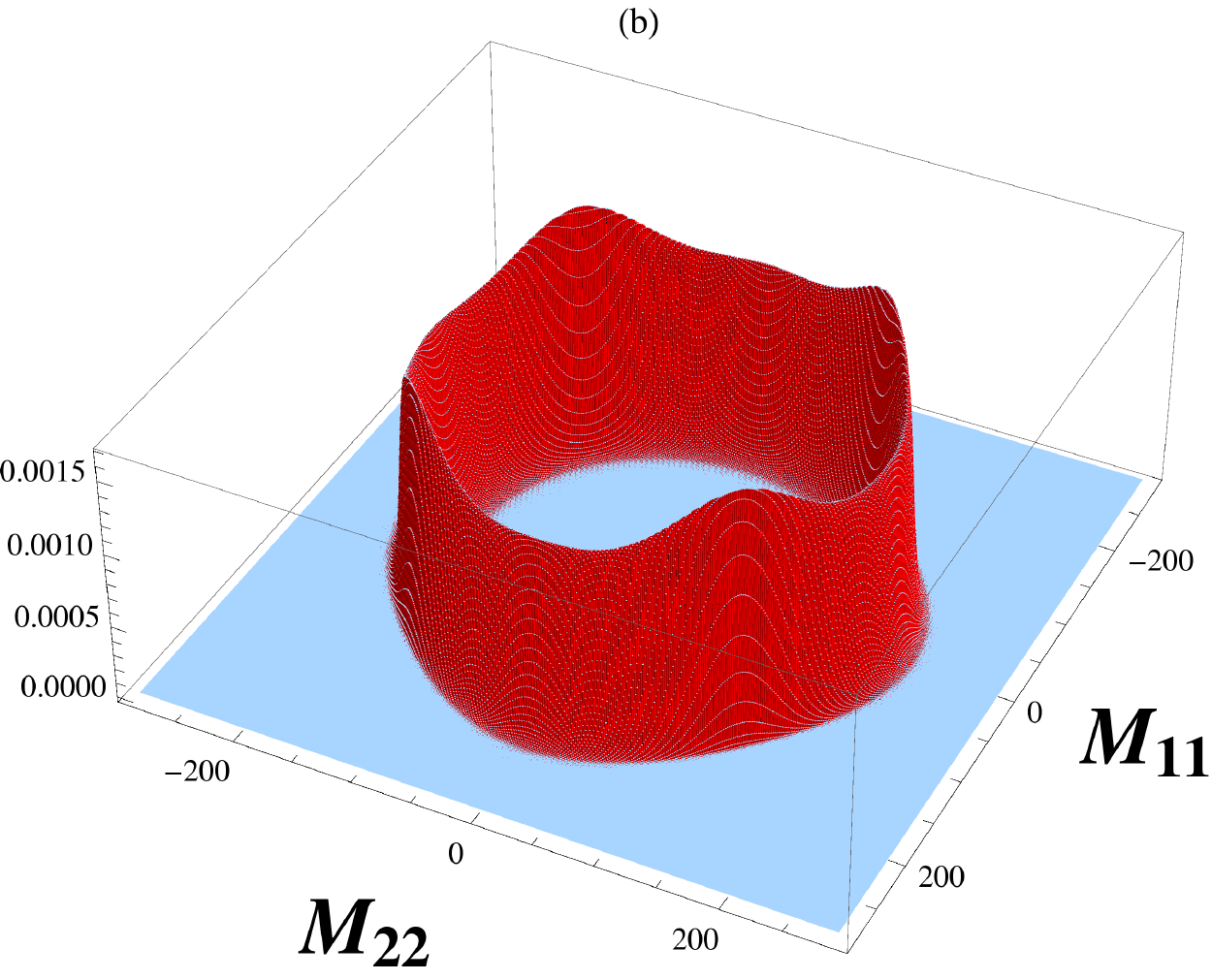} \\
\includegraphics[width=0.235\textwidth]{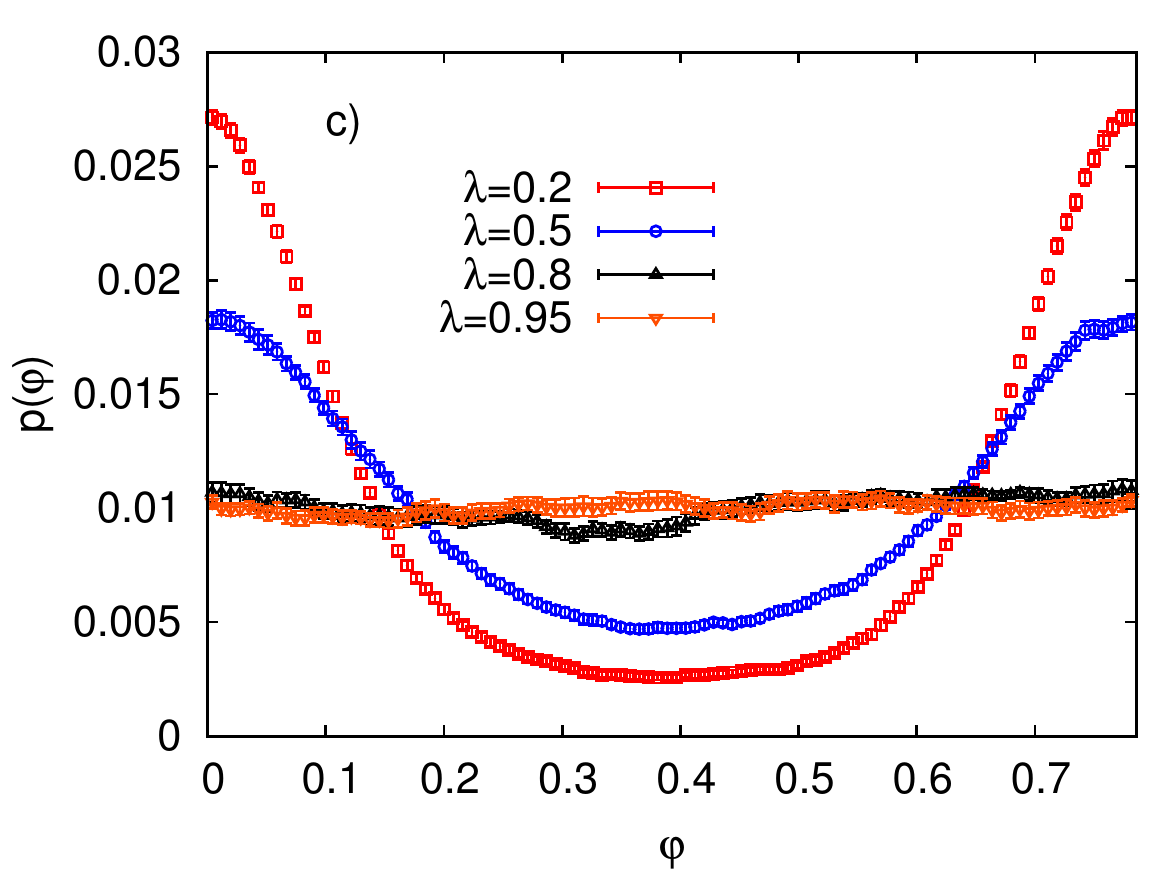}
\includegraphics[width=0.235\textwidth]{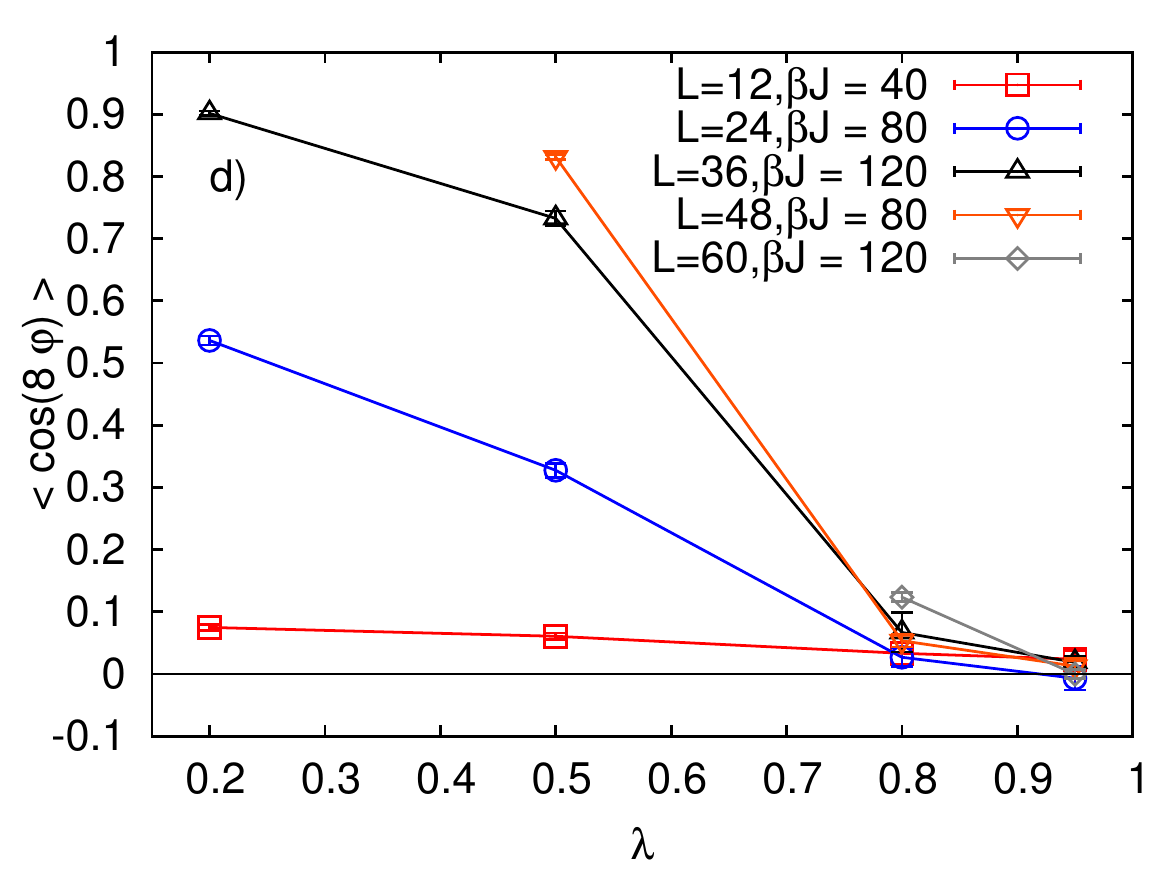}
\caption{[Color online] \textit{Probability distribution $p(M_{11},M_{22})$:
a) $\lambda = -0.5$, $\beta J = 50$, b) $\lambda = 0.8$, $\beta J = 100$,
$L_1 = L_2 = 24 a$. c) Probability distribution $p(\varphi)$ and d) 
$\langle \cos(8 \varphi) \rangle$.}}
\end{figure}

The angle $\varphi$ parametrizes the associated soft pseudo-Goldstone mode.
Since $\varphi$ and $\varphi + \pi$ are physically indistinguishable, only 
those states that are invariant against this shift belong to the physical 
Hilbert space. The corresponding low-energy effective theory is a $(2+1)$-d 
$\RP(1)$ model with the Euclidean Lagrangian
\begin{equation}
{\cal L} = \frac{\rho_t}{2} \partial_t \varphi \partial_t \varphi + 
\frac{\rho}{2} \partial_i \varphi \partial_i \varphi 
+ \kappa (\partial_i \partial_i \varphi)^2 + 
\delta \cos^2(4 \varphi).
\end{equation}
The effective theory predicts a finite-volume rotor spectrum (given simply by
$E_m = m^2/(2 \rho_t L_1 L_2)$ when assuming $\delta = 0)$ with 
$m = 0, \pm 2, \pm 4, \dots$. States with odd values of $m$ are excluded 
because they are not invariant against a shift of $\varphi$ by $\pi$. In fact,
the nine states with $m = 0, \pm 2, \pm 4, \pm 6, \pm 8$ have exactly the same 
(CT$_x$,CT$_y$) quantum numbers as the ones obtained by exact diagonalization 
(cf.\ Fig.3). The $\delta$-term explicitly breaks the emergent $SO(2)$ symmetry 
to a $\Z(8)$ subgroup and gives rise to a small Goldstone boson mass. The value 
of $\delta$ can be derived from a finite-size analysis of the probability 
distribution $p(\varphi)$ illustrated in Fig.4c. Inspired by \cite{Liu07}, we 
have determined the moment $\langle \cos(8 \varphi) \rangle = 
\int_{-\pi}^\pi d\varphi \ p(\varphi) \cos(8 \varphi)$ for different $\lambda$ 
(cf.\ Fig.4d).
We find that $\langle \cos(8 \varphi) \rangle > 0$ for $\lambda < 1$, indicating
that the system remains in the columnar phase all the way to the RK point. 
In mean field theory the most general quartic potential is given by
\begin{eqnarray}
&&V = \mu_1 O_1 + \mu_2 O_2 + \lambda_0 O_1 O_2 +
\sum_{i=1}^5 \lambda_i |O_i|^2, \nonumber \\
&&O_1 = M_{11}^2 +  M_{22}^2 + M_{12}^2 +  M_{21}^2, \nonumber \\
&&O_2 = M_{11} M_{12} - M_{11} M_{21} + M_{22} M_{12} + M_{22} M_{21}, \nonumber \\
&&O_3 = M_{11}^2 +  M_{22}^2 - M_{12}^2 -  M_{21}^2, \nonumber \\
&&O_4 = M_{11} M_{12} + M_{11} M_{21} - M_{22} M_{12} + M_{22} M_{21}, \nonumber \\
&&O_5 = M_{11} M_{22} + i M_{12} M_{21}.
\end{eqnarray}
A systematic analytic analysis of the minima of $V$ in the infinitesimal 
neighborhood of the staggered phase (which appears for $\lambda > 1$ and has 
$M_A = M_B = M_C = M_D = 0$, cf.\ Fig.1c) shall be presented elsewhere.

It is natural to define the dual field
$F_{\mu\nu}(x) = \frac{1}{\pi} \varepsilon_{\mu\nu\rho} \partial_\rho \varphi(x)$.
Vortices and half-vortices in the order parameter field manifest themselves as 
charges. The electric charge contained in a spatial region $\Omega$ is given by 
twice the vortex number
\begin{equation}
Q_\Omega = \int_\Omega d^2x \ \partial_i F_{0i} = 
\frac{1}{\pi} \int_{\partial \Omega} d\sigma_i \ 
\varepsilon_{ij} \partial_j \varphi \in \frac{\Z}{2}.
\end{equation}
Note that a charge 1 corresponds to a half-vortex, which is allowed because 
$\varphi$ and $\varphi + \pi$ are physically equivalent. The flux of $F_{\mu\nu}$
represents the conserved charges of the $U(1)^2$ center symmetry. At the RK 
point electric flux costs zero energy and condenses in the vacuum, thus giving 
rise to deconfinement even at zero temperature. In the effective theory, this 
implies that $\partial_i \varphi$ does not contribute to the energy, and hence 
that $\rho = \delta = 0$ at $\lambda = 1$. 

Figs.5a,b illustrate the energy distribution in an interface that separates two
distinct columnar phases with the columns oriented in the $y$-direction. By
complete wetting, this interface splits into two interfaces enclosing an 
intermediate columnar phase with the columns oriented in the $x$-direction. The
two interfaces separating the three columnar phases display plaquette order.
While the plaquette ordered region increases as one approaches the RK point, it
never becomes a stable bulk phase. 

Let us also investigate configurations with two external static charges $\pm 2$
relative to the staggered charge background, i.e.\ with $G_x = - (-1)^{x_1+x_2}$
at two positions separated by an odd number of lattice spacings. This violation
of the dimer covering constraint implies that three dimers touch each of those 
two sites (cf.\ Fig.2). The two external charges are confined by an electric 
flux string. Interestingly, the total flux 2 connecting the external charges 
$\pm 2$ fractionalizes into eight strands, each carrying electric flux 
$\frac{1}{4}$. The interior of the strands again displays plaquette order and 
represent interfaces separating distinct columnar phases whose columns are 
oriented in orthogonal directions. The energy distribution in the 
charge-anti-charge configuration is illustrated in Fig.5c. The energy of the 
string plays the role of a confining charge-anti-charge potential 
$V(r) \sim \sigma r$, which is shown in Fig.5d. Deep in the columnar phase, the 
string tension is given by $\sigma(\lambda = - 0.5) = 0.370(1) J/a$. With 
increasing $\lambda$, it is greatly reduced to 
$\sigma(\lambda = 0.5) = 0.057(1) J/a$,
$\sigma(\lambda = 0.8) = 0.029(1) J/a$, and it finally vanishes at the
deconfined RK point with $\sigma(\lambda = 1) = 0$. 
\begin{figure}[t]
\includegraphics[width=0.475\textwidth,angle=0]{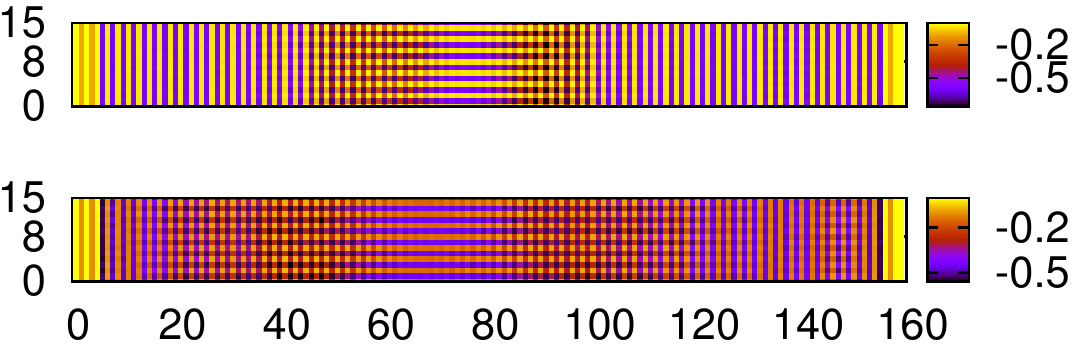} \\
\includegraphics[width=0.475\textwidth]{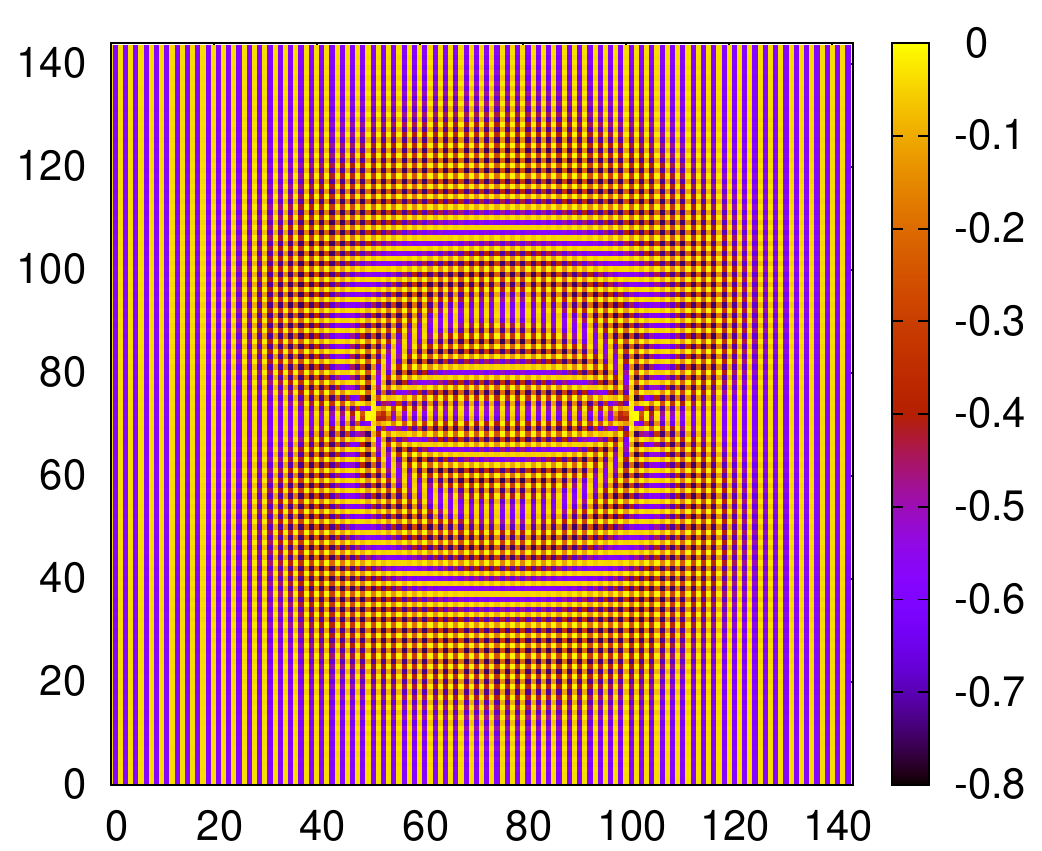} \\
\includegraphics[width=0.46\textwidth]{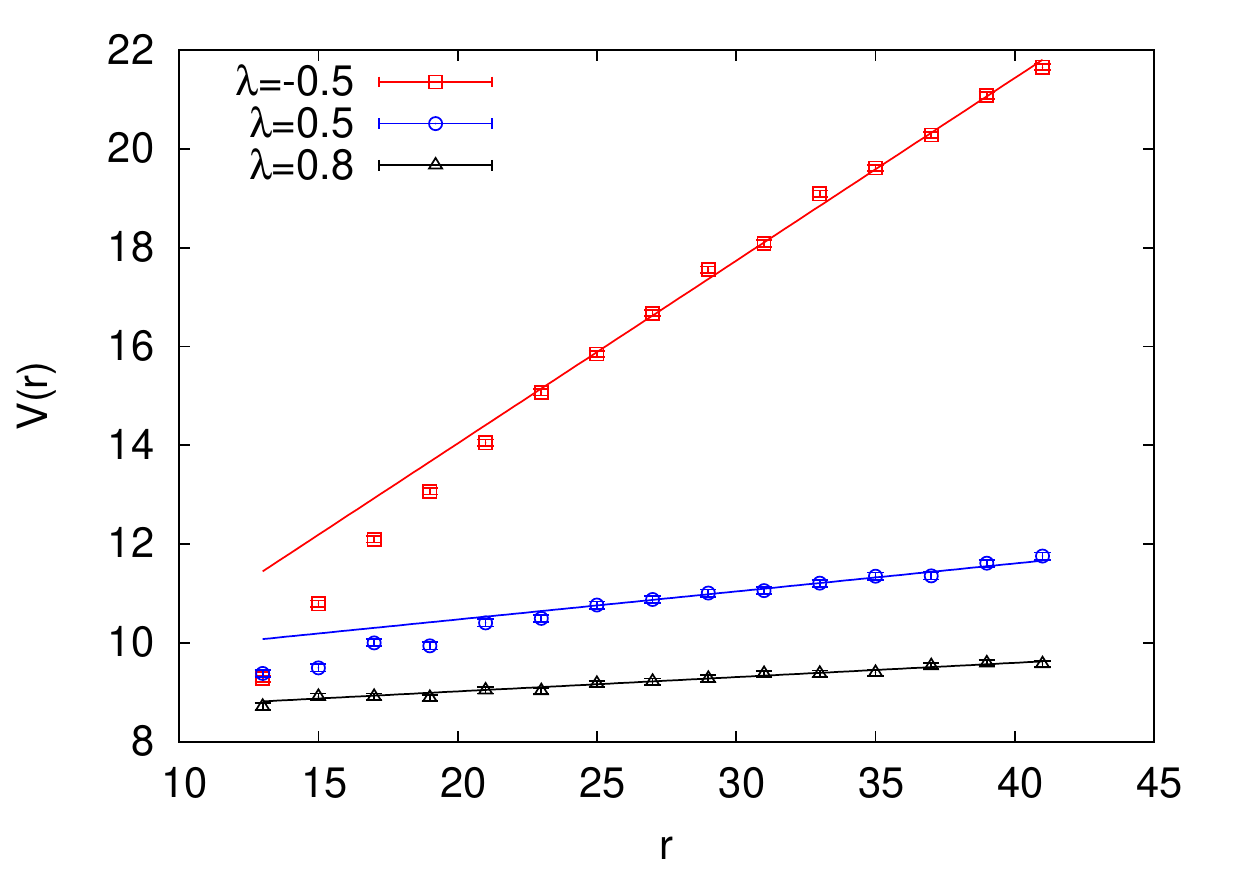}
\caption{[Color online] \textit{From top to bottom: Interfaces at a) 
$\lambda = - 0.5$ ($\beta J = 100$) and b) $\lambda = 0.7$ ($\beta J = 500$), 
on a $16 \times 160$ lattice. c) Energy density 
$-J \langle U_\Box + U_\Box^\dagger\rangle$ in the presence of 
two charges $\pm 2$ (separated by 49 lattice spacings) for $\lambda = - 0.2$, 
$\beta J = 72$, on a $144^2$ lattice. d) Potential between two static charges 
$\pm 2$ separated by a distance $r$ along a lattice axis, for 
$\lambda = - 0.5$, $0.5$, $0.8$.}}
\end{figure}

In conclusion, we found that the $(2+1)$-d quantum dimer model is in a columnar
phase for all values of the RK coupling $\lambda < 1$, without ever going into
a plaquette phase. At $\lambda = 1$, it is well known that electric fluxes 
condense in the vacuum, thus leading to deconfinement even at zero temperature. 
This corresponds to the spontaneous breakdown of the $U(1)^2$ center symmetry. 
In the confining columnar phase, the string connecting two external static 
charges separates into distinct strands, each carrying a fractionalized flux 
$\frac{1}{4}$. The interior of the flux strands consists of plaquette phase 
which does, however, not exist as a stable bulk phase. We have observed an 
approximate emergent $SO(2)$ symmetry with an associated pseudo-Goldstone 
boson.  At $\lambda = 1$ the Goldstone boson becomes exactly massless, and can 
be identified with a dual photon. Remarkably, as indicated by the rotor spectrum
in Fig.3, the pseudo-Goldstone mode, which
represents a soft phonon-like mode of the columnar valence bond solid, exists 
even far away from the RK point at $\lambda \approx 0$. It will be interesting 
to investigate whether this ``phonon'' mode persists in doped systems and 
whether it may be related to the formation of Cooper pairs in high-$T_c$ 
superconductors.

\newpage

UJW likes to thank the CTP at MIT, where this work was initiated, for 
hospitality during a sabbatical. D.\ B.\ acknowledges interesting discussions 
with A.\ L\"auchli. The research leading to these results has received funding 
from the Schweizerischer Na\-tio\-nal\-fonds and from the European Research 
Council under the European Union's Seventh Framework Programme 
(FP7/2007-2013)/ ERC grant agreement 339220.

\end{document}